\documentclass{article}%
\usepackage{amsfonts}
\usepackage{amsmath}
\usepackage{amssymb}
\usepackage{graphicx}%
\begin{document}

\title{Tunneling in mesoscopic junctions}
\author{H. O. Frota\\Departamento de F\'{\i}sica-ICE, Universidade Federal do Amazonas,\\{69077-000, Manaus-Am, Brazil}}
\maketitle

\begin{abstract}
We have applied the Numerical Renormalization Group method to study a
mesoscopic system consisting of two samples of metal separated by an
insulating barrier, with nanometer dimensions, which allows the tunnelling of
a single electron from one to the other side of the junction. The junction is
represented by a generalized orthodox model, taking into account the
electronic scattering interaction due the hole and the tunnelling electrons,
localized in the source and in the drain electrode, respectively. We have
calculated the static properties (charge transference, charge average,
quadractic charge average and specific heat) and the electric conductivity of
the junction, for the model parameters given by the tunneling matrices element
$V_{T}$, the barrier energy $U=e^{2}/2C$ (where $C$ is the capacitance of the
system) and by the electronic scattering potentials $V_{L(R)}$ acting on the
electrons of the left(right) electrode.

\end{abstract}

\section{\bigskip Introduction}

Studies of the tunnelling mesoscopic junctions in the present decade have
advanced thanks to the possibilities offered by the new manufacture
technologies of these structures, which are increasingly lesser, heaving
reached nanometric dimensions, resulting in capacitance lesser them
$10^{-15}F$, which permits to study the effect of the quantization of the
charge and electronic energy\cite{averin}.

The orthodox model for the tunnelling mesoscopic junctions consists of two
metal layers intercalated by a fine insulating layer. From the classic point
of view there is no tunnelling current through the junction, due to the
impediment of electrons to tunnel through the potential barrier represented by
the insulator. In this condition the junction behaves as a capacitor with $C$
capacitance . By applying an external potential $V_{ext}$ to the junction, it
is loaded with a charge $Q=CV_{ext}$. This charge, which can vary
continuously, is originated by the displacement of electrons in the electrodes
in relation to the\ positive ions of metals. Even lowermost, $V_{ext}$ can
produce small displacements of electrons, originating a small charge in the
electrodes. The interaction between the charges of these electrodes can be
represented by the energy of the capacitor that is given by $E_{c}=Q^{2}/2C$.

Taking in consideration the quantum effect, when the layer of insulator,
intercalated enters the metal layers, will be very fine, it is possible that
an electron tunnels from a side to the other side of the junction, through the
insulator layer. In this case the process of electronic transport involves
discrete charge of only one electron ($e$), in contrast of what occurs with
the charges of the electrodes that are continuous. As the electrostatic energy
associated with the tunnelling of an electron is $U=e^{2}/2C$, for great
values of $C$ this energy is masked by the thermal energy $k_{B}T$. However,
with the advance of the nanostructure manufacture techniques, it is possible
to construct junctions with small capacitance, so that the electrostatic
energy $U>k_{B}T$. In these conditions, the effect of the charge transference
can be observed. Then, considering that the tunnelling conductance $\Gamma$
also is low, much smaller then the inverse of the quantum resistance of the
junction, $\Gamma\ll R_{Q}^{-1}$, where $R_{Q}=\pi\Gamma/(2e^{2}%
)\simeq6.5K\Omega$, the tunnelling of a single electron can hinder the
tunnelling of the following electrons, giving \ rise to what is known as
Coulomb blockade \cite{averin,grabert,kastner}, that has excited great
theoretical and experimental interest. Recent works have shown that, even at
energies below the energy of capacitor $E_{c}$, there is tunnelling due to
charge fluctuation in the junction \cite{flensberg,frota2,guinea1}.

In the present work we have generalized the orthodox model, considering that
the charge transference from a side (source) of the junction to the other
(drain) creates scattering potentials in the electrodes, originated by the
hole left in the source and by the electron that has tunnelled to the drain.
The sudden creation of these localized potentials creates electron-hole pairs
excitation at the Fermi level, modifying the behavior of the conductivity of
the junction. The tunnelling process is in thermal equilibrium when the excess
charge in each junction, originated from the tunnelling, is balanced by the
phase shift of the conduction electrons, according to the Fridel sum rule.
This problem has been treated similarly to the problem related to X-ray
threshold singularity \cite{guinea2,drewes}, where the many particles states,
after the tunnelling, are projected on the many particles states before the
tunnelling, what can induce X-ray infra-red divergence in the conductivity
\cite{drewes}.

Other forms of nanostructures have deserved increasing interest from
researchers, as the structure formed by two metal layers intercalated by an
insulator layer, inside of which there is a metallic electrode, that usually
receives the name of island \cite{fulton}. The process of charge transport
from a metal (source) to the other (drain) occurs by quantum tunnelling
through the island, changing its charge by $e$. The effect of the tunnelling
of an electron through the island can be enhanced if the dimension of the
island is very small, what confers to it a very reduced capacitance and,
consequently, a great variation of its potential. This high potential, that
produces an electrostatic energy higher then the thermal energy, reduces the
probability of tunnelling through the island, as it has been observed in
granular metallic materials in the beginning of the studies of the tunnelling
junctions \cite{gorter,neugebauer,lambe}, where the electronic tunnelling in
low voltages is inhibited when the electrostatic energy of an electron of a
grain is greater that the thermal energy. With the development of the
manufacture of nanostructure technology it is possible to construct metallic
islands with known geometry, separated by tunnelling barriers, that can
adequately control the tunnelling of a single electron through the junction,
that occurs by the jump of this electron from one to another island
\cite{fulton,schon,geerligs,pothier}.

Another geometry of nanostruture that has been studied is constituted by a
metal substratum over which a layer of insulator and another one of
semiconductor are deposited successively \cite{meirav}, with a pair of
electrodes build parallel on the layer of the semiconductor. A positive
voltage is applied to the metal substratum, so that the free electrons of the
semiconductor are confined to the superior surface of the insulator. This
confinement forms a bidimensional free electron gas in the interface
semiconductor-insulator. By applying a negative voltage to the pair of
electrodes on the layer of the semiconductor, the movement of the electrons is
limited throughout a canal in the parallel direction to the electrodes, thus
forming a free unidimensional electron gas. Two saliences constructed in the
electrodes have the purpose of creating potential barriers that confine
electrons between them, producing discrete energy levels in this region,
reason why these devices are called \textit{quantum points} or
\textit{artificial atoms }%
\cite{kastner1,kastner2,meir,wilkins,ding,flensberg2,ralph,furusaki1,furusaki2}%
.

In this work, using the Numerical Renormalization Group (NRG) formalism, we
have developed an accurate numerical calculation for the generalized orthodox
model for mesoscopic junctions, taking into account the electronic scattering
\cite{guinea2,drewes}. We have initiated by presenting the Hamiltonian of the
Model in section I, where the properties that will be studied have been
defined; in section II we have developed the formalism adopted for the
diagonalization of the Hamiltonian, the calculation of the static properties
and the electric conductivity; the section III consists in the central point
of this work, where the results obtained for the calculations of the static
properties (charge transference, charge average, quadratic charge average and
the specific heat) and of the electric conductivity of the model are
presented; in section IV we have presented the pertinent conclusions to this
work and, finally, in the appendix we have developed a perturbation theory in
the tunnelling matrices element $V_{T}$ for a particular case of the studied
model, whose results are used to verify of the precision of the numerical
calculation of the general model.

\section{The model}

The tunnelling junction studied is formed by two metals which are
separated\ by an insulator layer, whose dimensions are nanometrics. The small
thickness of the insulator allows the electrons to tunnel from a metal to the
other, changing the charge pattern of the junction and giving origin to a
tunnelling current. Usually the the Coulomb blockade theory assumes that the
conduction electrons, before the tunnelling event, are in an equilibrium state
and, after the tunnelling, these electrons enters immediately into a new
equilibrium state \cite{averin}. This means that the wave functions of the
electronic states next to the Fermi level are not changed with the variation
of the charge junction. Only a small displacement in the energy levels of the
electrons occurs next to the Fermi level. This theory, therefore, fails to
take in consideration the transient effect in the charging patterns between
the two states of equilibrium.

To account for the transient effect, some authors \cite{guinea2,drewes,ueda2}
have considered that the tunnelling event occurs in two stages, with different
time scales. In the first stage of the electron tunnelling through the
junction, the charge created in the surface of the metal due to the
transference of an electron from one side to the other side of the junction,
enter in equilibrium in a time of the order of the inverse of the plasma
frequency of the metal, whose corresponding energy is of approximately 1eV. In
the second stage, the electrons in the Fermi surface, which are associated to
a response time much longer then the correspondent to the first stage, feel
the change in the charge states of the junction as if this has occurred
quickly and in a localized form. This difference in the time scale, associated
with the nature of the localized potential quickly created in the surface of
the metal, generates many particles interactions, giving origin to the
electron-hole pairs excitations in the Fermi level. This effect is similar to
the Mahan-Nozi\`{e}res-Dominicis effect \cite{mahan,nozieres} that occurs in
the absorption and photoemission of X-rays in metals. In this in case an
electron is ejected from a deep atomic level of a metal, changing the charge
of an atom. The ion quickly created by the emission of the electron by the
ray-X interacts in the final state with conduction electrons, resulting in a
divergence in the X-ray absorption and photoemission spectrum.

In this section we have presented the Hamiltonian of the model for a
nanostructure metal-insulator-metal, considering the electrostatic screen
originated from the tunnelling of the electron through the junction. The model
Hamiltonian is written as%

\begin{equation}
H=H_{L}+H_{R}+H_{T}+H_{Q}+H_{V}\text{.} \label{hmodelo}%
\end{equation}

The terms $H_{L}$ and $H_{R}$ represent the left and right electrodes
conduction Hamiltonians, respectively, given by%

\begin{align}
H_{L}  &  =\sum_{k}\epsilon_{k}c_{k}^{\dagger}c_{k}\label{bandal}\\
H_{R}  &  =\sum_{q}\epsilon_{q}d_{q}^{\dagger}d_{q} \label{bandar}%
\end{align}
where the operators $c_{k}^{\dagger}$($c_{k}$) and $d_{q}^{\dagger}$($d_{q} $)
create (annihilate) electrons in the respective conduction band and obey the
following anticommutation relations:%

\begin{equation}%
\begin{array}
[c]{lllll}%
\left\{  c_{k}^{\dagger},c_{k^{,}}\right\}  & = & \delta_{kk^{,}} &  & \\
\left\{  c_{k},c_{k^{,}}\right\}  & = & \left\{  c_{k}^{\dagger},c_{k^{,}%
}^{\dagger}\right\}  & = & 0\\
\left\{  d_{q}^{\dagger},d_{q^{,}}\right\}  & = & \delta_{qq^{,}} &  & \\
\left\{  d_{q},d_{q^{,}}\right\}  & = & \left\{  d_{q}^{\dagger},d_{q^{,}%
}^{\dagger}\right\}  & = & 0\\
\left\{  c_{k},d_{q}\right\}  & = & 0 &  &
\end{array}
. \label{anticom}%
\end{equation}
The last of the proprieties indicates the independence of the operators of the
two conduction bands. The term $H_{T}$ describes the tunnelling Hamiltonian%

\begin{equation}
H_{T}=%
{\displaystyle\sum_{kq}}
\left(  V_{T}c_{k}^{\dagger}d_{q}+V_{T}^{\ast}d_{q}^{\dagger}c_{k}\right)  ,
\label{htunel}%
\end{equation}
which allows the electrons be annihilate in the left side conduction band and
be create in the right side conduction band, and vice versa, and $V_{T}$\ is
the tunnelling matrices element, which is taken as independent of the moment
for energies very close to de Fermi energy. The electrostatic junction energy
is represented by the term $H_{Q}$,%

\begin{equation}
H_{Q}=\frac{Q^{2}}{2C} \label{hq}%
\end{equation}
where $C$ is the capacitance, $Q$ is the junction charge, given by%

\begin{align}
Q  &  =\frac{e}{2}\left(  N_{L}-N_{R}\right) \label{q}\\
N_{L}  &  =\sum_{k}c_{k}^{\dagger}c_{k}\label{nl}\\
N_{R}  &  =\sum_{q}d_{q}^{\dagger}d_{q}, \label{nr}%
\end{align}
and $e$ is the elementary charge.

The last term of the Hamiltonian given by the Eq.(\ref{hmodelo}) corresponds
to the free electron scattering by the charges localized in the junctions,
written as%

\begin{equation}
H_{V}=V_{L}\frac{\left(  N_{L}-N_{R}\right)  }{2}%
{\displaystyle\sum_{kk^{,}}}
c_{k}^{\dagger}c_{k^{,}}+V_{R}\frac{\left(  N_{L}-N_{R}\right)  }{2}%
{\displaystyle\sum_{qq^{,}}}
d_{q}^{\dagger}d_{q^{,}}\text{,} \label{hv}%
\end{equation}
where $V_{L}$ and $V_{R}$ represent the changing in the potential of the
metals in the left and right side of the junction, respectively, when an
electron tunnels trough the junction.

\bigskip

\section{The Method}

\bigskip

\bigskip The Numerical Renormalization Group formalism used in this paper
follows the method proposed originally by K. G. Wilson\cite{wilson1} for the
solution of the static proprieties of the Kondo model, and applied by
Krishna-murthy, Wilkins and Wilson\cite{krishna1,krishna2} to calculate these
proprieties in the Anderson model, later generalized by Frota and
Oliveira\cite{frota1} to obtain the dynamics proprieties of these models.

The first study using the Wilson renormalization group method for
nanostructures was carried through by Frota and Flensberg \cite{frota2},
having determined the fundamental state and transports proprieties for these
structures, represented by the usual model, which was not taken in
consideration that when an electron tunnels through a junction, the conduction
electrons quickly readjust its potential, created in the surface of the
electrode, producing many particle effects with the creation of electron-hole
pairs excitations. In the present work we have introduced this new ingredient
and we have analyzed its effect in the fundamental state and transports properties.

The method consists in the logarithmic discretization of the Hamiltonian of
the conduction bands of the two electrodes, and in the determination of a new
finite basis in which the Hamiltonian of the model, given by the
Eq.(\ref{hmodelo}), is written. After the iterative diagonalization, we obtain
its eingenvalues and eigenvectors, calculate the charge average and the
quadratic charge average of the junction, specific heat and, with the Kubo
formula, we obtain the electric conductivity. The conduction bands are divided
into logarithmic intervals $\pm D\Lambda^{-j-z}$ $(j=1,2,...$; $\Lambda>1;$
$z$ is a continuous parameter$)$, and converted to a ``hopping'' Hamiltonian
\cite{krishna1} given by%

\begin{align}
H_{L}^{N}  &  =\frac{\left(  1+\Lambda^{-1}\right)  D}{2}%
{\displaystyle\sum_{n=0}^{N-1}}
\varepsilon_{nz}\left(  f_{nz}^{\dagger}f_{(n+1)z}+f_{(n+1)z}^{\dagger}%
f_{nz}\right) \label{hltrunc}\\
H_{R}^{N}  &  =\frac{\left(  1+\Lambda^{-1}\right)  D}{2}%
{\displaystyle\sum_{n=0}^{N-1}}
\varepsilon_{nz}\left(  g_{nz}^{\dagger}g_{(n+1)z}+g_{(n+1)z}^{\dagger}%
g_{nz}\right)  \text{,} \label{hrtrunc}%
\end{align}
where $\varepsilon_{nz}$ is obtained numerically. In this basis the electrodes
are coupling solely via the first terms of the chains, defined as%

\begin{align}
f_{0z}  &  =\frac{1}{\sqrt{2}}\sum_{k}c_{k}\label{f0z}\\
g_{0z}  &  =\frac{1}{\sqrt{2}}\sum_{q}d_{q}\text{.} \label{g0z}%
\end{align}

In the basis $\{f_{nz},g_{nz}\}$, the tunnelling Hamiltonian $H_{T},$ the
electrostatic junction energy $H_{Q}$ and \ the scattering Hamiltonian
$H_{V}^{N}$ are given by%

\begin{align}
H_{T}^{N}  &  =2\left(  V_{T}f_{0z}^{\dagger}g_{0z}+V_{T}^{\ast}%
g_{0z}^{\dagger}f_{0z}\right) \label{htf0g0}\\
H_{Q}^{N}  &  =\frac{U}{4}\left(  \sum_{n=0}^{N}f_{nz}^{\dagger}f_{nz}%
-\sum_{n=0}^{N}g_{nz}^{\dagger}g_{nz}\right)  ^{2}\label{hqn}\\
H_{V}^{N}  &  =\left(  \sum_{n=0}^{N}f_{nz}^{\dagger}f_{nz}-\sum_{n=0}%
^{N}g_{nz}^{\dagger}g_{nz}\right)  \left(  V_{L}f_{0z}^{\dagger}f_{0z}%
+V_{R}g_{0z}^{\dagger}g_{0z}\right)  \text{.} \label{hvn}%
\end{align}

\strut Rescaling by the factor $2\Lambda^{-(N-1)/2}/[(1+\Lambda^{-1})D]$ the
Hamiltonians given by Eqs.(\ref{hltrunc},\ref{hrtrunc},\ref{htf0g0},\ref{hqn}
and \ref{hvn}), the diagonalization takes place in the reduced Hamiltonian
$H_{N}$ that is written as%

\begin{align}
H_{N}  &  =\Lambda^{(N-1)/2}\left\{
{\displaystyle\sum_{n=0}^{N-1}}
\varepsilon_{nz}\left(  f_{nz}^{\dagger}f_{(n+1)z}+f_{(n+1)z}^{\dagger}%
f_{nz}\right)  \right. \nonumber\\
&  +%
{\displaystyle\sum_{n=0}^{N-1}}
\varepsilon_{nz}\left(  g_{nz}^{\dagger}g_{(n+1)z}+g_{(n+1)z}^{\dagger}%
g_{nz}\right) \nonumber\\
&  +\widetilde{V}_{T}\left(  f_{0z}^{\dagger}g_{0z}+g_{0z}^{\dagger}%
f_{0z}\right) \label{hn}\\
&  +\frac{\widetilde{U}}{4}\left(  \sum_{n=0}^{N}f_{nz}^{\dagger}f_{nz}%
-\sum_{n=0}^{N}g_{nz}^{\dagger}g_{nz}\right)  ^{2}\nonumber\\
&  +\left.  \frac{1}{2}\left(  \sum_{n=0}^{N}f_{nz}^{\dagger}f_{nz}-\sum
_{n=0}^{N}g_{nz}^{\dagger}g_{nz}\right)  \left(  \widetilde{V}_{L}%
f_{0z}^{\dagger}f_{0z}+\widetilde{V}_{R}g_{0z}^{\dagger}g_{0z}\right)
\right\}  ,\nonumber
\end{align}
where%

\begin{align}
\widetilde{V}_{T}  &  =\frac{4V_{T}}{\left(  1+\Lambda^{-1}\right)
D}\label{ttil}\\
\widetilde{U}  &  =\frac{2U}{\left(  1+\Lambda^{-1}\right)  D}\label{util}\\
\widetilde{V}_{L}  &  =\frac{4V_{L}}{\left(  1+\Lambda^{-1}\right)
D}\label{vltil}\\
\widetilde{V}_{R}  &  =\frac{4V_{R}}{\left(  1+\Lambda^{-1}\right)  D}.
\label{vrtil}%
\end{align}
The scale factor $\Lambda^{(N-1)/2}$ in the reduced Hamiltonian $H_{N}$ is
introduced to assure that the smallest eigenvalue is of unity order. The
discrete approximation to the Hamiltonian of the model is given as the limit%

\begin{equation}
H=\lim_{N\rightarrow\infty}\frac{1}{2}(1+\Lambda^{-1})D\Lambda^{-(N-1)/2}H_{N}
\label{hcontinuous}%
\end{equation}

The diagonalization of the Hamiltonian $H_{N}$, represented by the Eq.
(\ref{hn}), is carried out iteratively, using a recurrence relation which
allows to calculate the eigenstates of the iteration (N+1) from the
eigenstates of the iteration N. To obtain this recurrence relation, the charge
operator given by Eq.(\ref{q}) in the basis \{$f_{n},g_{n}$\} is writtten as,%

\begin{equation}
Q_{N}\equiv\frac{1}{2}\left(  \sum_{n=0}^{N}f_{nz}^{\dagger}f_{nz}-\sum
_{n=0}^{N}g_{nz}^{\dagger}g_{nz}\right)  \label{qn}%
\end{equation}
which commutes with the operator $H_{N}$, since $[Q_{N},H_{N}]=0$. The
definition of the charge operator is important to the numerical
diagonalization process of the Hamiltonian $H_{N}$ because, as $Q_{N}$ and
$H_{N}$ commute, the associated matrices can be written as blocks of matrices
that can be diagonalized independently, lowering the computational time. From
the definition of $Q_{N}$, the recurrence relation for the Hamiltonian
$H_{N\text{ }}$is written as%

\begin{align}
H_{N+1}  &  =\sqrt{\Lambda}H_{N}+\Lambda^{N/2}I_{N}+2\Lambda^{N/2}%
\widetilde{U}Q_{N}q_{N+1}+\Lambda^{N/2}\widetilde{U}\left(  q_{N+1}\right)
^{2}\nonumber\\
&  +\Lambda^{N/2}q_{N+1}\left(  \widetilde{V}_{L}f_{0z}^{\dagger}%
f_{0z}+\widetilde{V}_{R}g_{0z}^{\dagger}g_{0z}\right)  , \label{hnmais1}%
\end{align}
where the iterative operator $I_{N}$, which couples the iteration ($N$) with
the iteration ($N+1$), is given by%

\begin{equation}
I_{N}=\varepsilon_{Nz}\left[  f_{Nz}^{\dagger}f_{(N+1)z}+f_{(N+1)z}^{\dagger
}f_{Nz}+g_{Nz}^{\dagger}g_{(N+1)z}+g_{(N+1)z}^{\dagger}g_{Nz}\right]  .
\label{iteration}%
\end{equation}
and $q_{N}$ is defined as%

\begin{equation}
q_{N}=\frac{1}{2}\left(  f_{Nz}^{\dagger}f_{Nz}-g_{Nz}^{\dagger}g_{Nz}\right)
\label{qnminus}%
\end{equation}

From Eqs. (\ref{qn} and \ref{qnminus}), the recurrence relation for $Q_{N}$
and $Q_{N}^{2}$ is written as%

\begin{equation}
Q_{N+1}=Q_{N}+q_{N+1} \label{qnmais1}%
\end{equation}

\begin{equation}
\left[  Q_{N+1}\right]  ^{2}=\left[  Q_{N}\right]  ^{2}+2\left[  Q_{N}\right]
q_{N+1}+\left[  q_{N+1}\right]  ^{2}, \label{qnmais12}%
\end{equation}
which will be useful for the calculation of the charge average and the
quadratic charge \ \ average.

The Eq.(\ref{hnmais1}) consists in the core of the renormalization group
method used in the present work. It permits that, once the eigenvalues and
eigenvectors of the Hamiltonian $H_{N}$ are known, the eignvalues and the
eigenvectors of the Hamiltonian $H_{N+1}$ can be determined.

For a given iteration $N$, the eigenstates consist in the following Fermi
operators: $f_{0z},$ $f_{1z},...,f_{Nz}$ and $g_{0z},g_{1z},...,g_{Nz}$. Each
one of these operators can assume two states, occupied and empty, so that for
the iteration $N$ the number of states is given by $2^{2(N+1)}$ and the
matrices associate to the Hamiltonian $H_{N}$ in the basis $\{f_{nz},g_{nz}\}$
is of the order $2^{2(N+1)}\times2^{2(N+1)}$. For tipical values of $N$ used
in the present work ($N=20$), the matrices that will be diagonalized are of
the order $2^{42}\times2^{42}$, which makes the numerical diagonalization
impossible. To by-pass these difficulties were carried out, two procedures
which are reported as follows: 1) As the operator $Q_{N}$ commutes with the
Hamiltonian $H_{N}$, then the eigenstates of $H_{N}$ also are eigenstates of
$Q_{N}$, which means that we can diagonalize $H_{N}$ in independent subspaces
of $Q_{N}$. Thus, the matrices associated with the Hamiltonian $H_{N}$ can be
written in diagonal blocks that are diagonalized separately, which reduces the
dimensions of the matrices to be diagonalized. 2) Even if diagonalizing in
subspaces of same charge, the matrices continue with dimensions that makes the
numerical diagonalization process inadequate. However, as in the calculation
of the static and dynamic proprieties the energies are close to the Fermi
energy, the procedure consists in generating the basis for the diagonalization
of the iteration ($N+1$) taking into account only the lower energy eigenstates
of the iteration ($N$), according with the references \cite{wilson1} and
\cite{krishna1}.

The iterative process begins with the diagonalization \ of the Hamiltonian
$H_{0}$, written as%

\begin{align}
H_{0}  &  =\frac{1}{\sqrt{\Lambda}}\left\{  \widetilde{V}_{T}\left(
f_{0z}^{\dagger}g_{0z}+g_{0z}^{\dagger}f_{0z}\right)  +\widetilde{U}\left[
\frac{1}{2}\left(  f_{0z}^{\dagger}f_{0z}-g_{0z}^{\dagger}g_{0z}\right)
\right]  ^{2}\right. \label{h0}\\
&  +\left.  \frac{1}{2}\left(  f_{0z}^{\dagger}f_{0z}-g_{0z}^{\dagger}%
g_{0z}\right)  \left(  \widetilde{V}_{L}f_{0z}^{\dagger}f_{0z}+\widetilde
{V}_{R}g_{0z}^{\dagger}g_{0z}\right)  \right\}  ,\nonumber
\end{align}
where only the operators $f_{0z}$ and $g_{0z}$ appear. \ In the second step,
the operators $f_{1z}$ and $g_{1z}$ are added and the Hamiltonian%

\begin{equation}
H_{1}=\sqrt{\Lambda}H_{0}+I_{0}+2\widetilde{U}Q_{0}q_{1}+\widetilde{U}%
q_{1}^{2}+q_{1}\left(  \widetilde{V}_{L}f_{0z}^{\dagger}f_{0z}+\widetilde
{V}_{R}g_{0z}^{\dagger}g_{0z}\right)  ,
\end{equation}
is diagonalized in relation to a basis formed by the eigenstates of the
Hamiltonian in the first step ($H_{0}$) and the states that are constructed by
the operators $f_{1z}^{\dagger}$, $g_{1z}^{\dagger}$ and $f_{1z}^{\dagger
}g_{1z}^{\dagger}$ applied to the eigenstates of $H_{0}$. Following this
procedure, the Hamiltonian $H_{N+1}$ is diagonalized in relation to a basis
formed by the eigenstates of $H_{N}$ and the states given by the operator
$f_{\left(  N+1\right)  z}^{\dagger}$, $g_{\left(  N+1\right)  z}^{\dagger}$
and $f_{\left(  N+1\right)  z}^{\dagger}g_{\left(  N+1\right)  z}^{\dagger}$
applied to the eigenstates of $H_{N}$.

For each iteration the charge average $\left\langle Q_{N}\right\rangle $, the
quadratic charge average $\left\langle Q_{N}^{2}\right\rangle $, the specific
heat $C_{V}$, the charge transference $\tau$ and the electrical current
$\sigma(\omega)$ are calculated, in relation to the basis \{$f_{0z}%
,g_{0z},f_{1z},g_{1z},...,f_{Nz},g_{Nz},$\}, as follows,%

\begin{align}
\left\langle Q_{N}\right\rangle  &  =\frac{\text{Tr}Q_{N}\exp\left(
-\bar{\beta}H_{N}\right)  }{\text{Tr}\exp\left(  -\bar{\beta}H_{N}\right)
}\label{qnmedio}\\
\left\langle Q_{N}^{2}\right\rangle  &  =\frac{\text{Tr}Q_{N}^{2}\exp\left(
-\bar{\beta}H_{N}\right)  }{\text{Tr}\exp\left(  -\bar{\beta}H_{N}\right)
}\label{qnmedio2}\\
\frac{C_{V}}{k_{B}}  &  =\overline{\beta}^{2}\left[  \frac{\text{Tr}\left(
H_{N}\right)  ^{2}\exp\left(  -\bar{\beta}H_{N}\right)  }{\text{Tr}\exp\left(
-\bar{\beta}H_{N}\right)  }-\left(  \frac{\text{Tr}H_{N}\exp\left(
-\bar{\beta}H_{N}\right)  }{\text{Tr}\exp\left(  -\bar{\beta}H_{N}\right)
}\right)  ^{2}\right]  \label{cv}%
\end{align}
where the Boltzmann factor is written in terms of the reduced Hamiltonian
$H_{N}$ as%

\begin{equation}
\exp(-\beta H)=\exp\left[  -\bar{\beta}H_{N}\right]
\end{equation}
where $\beta=$1/k$_{B}T$, k$_{B}$ is the Boltzmann constant, $T$ is the
temperature and $\bar{\beta}$ a constant that is defined according with
reference \cite{krishna1},%

\begin{equation}
\overline{\beta}=-\beta\left[  \frac{1}{2}\left(  1+\Lambda^{-1}\right)
\Lambda^{-(N-1)/2}\right]
\end{equation}
Then, taking a fixed $\bar{\beta}<1$ \cite{krishna1}, the temperature $T_{N}$,
corresponding to the iteration $N$, is given by%

\begin{equation}
T_{N}=\frac{D}{k_{B}\overline{\beta}}\left[  \frac{1}{2}\left(  1+\Lambda
^{-1}\right)  \Lambda^{-(N-1)/2}\right]  .
\end{equation}

The term $H_{T}$ of the Hamiltonian $H$ allows the tunnelling through the
junction, given by the charge transference operator $\tau=\sum\nolimits_{kq}%
c_{k}^{\dagger}d_{q}$, which is written in the basis $\left\{  f_{nz}%
,g_{nz}\right\}  $ as
\begin{equation}
\tau=2f_{0z}^{\dagger}g_{0z}.
\end{equation}
The charge transference through the junction is given by the thermal average
of the operator $\tau$,%

\begin{equation}
\left\langle \tau\right\rangle =\frac{\text{Tr}\left\langle 2f_{0z}^{\dagger
}g_{0z}\right\rangle \exp\left(  -\bar{\beta}H_{N}\right)  }{\text{Tr}%
\exp\left(  -\bar{\beta}H_{N}\right)  }. \label{tz}%
\end{equation}

The electrical current operator through the junction, $I$ , is written as a
function of the\ time rate of the charge $Q$ as
\begin{align}
I  &  =\frac{dQ}{dt}\nonumber\\
&  =\frac{e}{2}\frac{d(N_{L}-N_{R})}{dt}. \label{i1}%
\end{align}
The above time derivative is obtained from the commutation of the operator
$N_{L}-N_{R}$ with the Hamiltonian $H$ given by Eq.(\ref{hmodelo}). As
$N_{L}-N_{R}$ commutes with $H_{L}$, $H_{R}$, $H_{Q}$ and $H_{V}$, then%

\[
\frac{d(N_{L}-N_{R})}{dt}=\frac{i}{\hslash}\left[  H_{L},(N_{L}-N_{R})\right]
,
\]
and the electrical current operator is written in terms of the operators
$c_{k}$ and $d_{q}$ as%

\begin{equation}
I=\frac{ei}{2\hslash}%
{\displaystyle\sum_{k,q}}
\left(  V_{T}c_{k}^{\dagger}d_{q}-V_{T}^{\ast}d_{q}^{\dagger}c_{k}\right)  .
\label{i2}%
\end{equation}

The electrical conductivity through the junction is obtained from the Kubo
formula\cite{mahan}:%

\begin{equation}
\sigma(\omega)=\frac{\pi}{\omega}\sum_{F}\left|  \left\langle F\right|
I\left|  \Omega\right\rangle \right|  ^{2}\delta\left(  E^{F}-E^{\Omega
}-\omega\right)  , \label{condut}%
\end{equation}
where $\omega$ is the frequency, $\left|  \Omega\right\rangle $ and $\left|
F\right\rangle $ are the fundamental and the final many particles states of
the Hamiltonian $H$, with eigenvalues $E^{\Omega}$ and $E^{F}$, respectively.

For NRG purpose, the operator $I$, given by Eq.(\ref{i2}), is written in terms
of the operators $f_{0z}$ and $g_{0z}$ (Eq. (\ref{f0z} and \ref{g0z}), respectively),%

\begin{equation}
I(z)=\frac{ei}{\hslash}\left(  V_{T}f_{0z}^{\dagger}g_{0z}-V_{T}^{\ast}%
g_{0z}^{\dagger}f_{0z}\right)  , \label{iz}%
\end{equation}
and $\sigma(\omega)$ (Eq.(\ref{condut})) is written in terms of the
eigenstates of the reduced Hamiltonian $H_{N}$ (Eq.(\ref{hn})), as%

\begin{equation}
\sigma(\omega_{N},z)=\frac{\pi}{\omega_{N}}\frac{%
{\displaystyle\sum\limits_{F}}
\left|  \left\langle F(z)\right|  I(z)\left|  \Omega(z)\right\rangle \right|
^{2}}{\left(  \frac{1+\Lambda^{-1}}{2}\Lambda^{-(N-1)/2}\right)  ^{2}}%
\delta\left(  E_{N}^{F}(z)-E_{N}^{\Omega}(z)-\omega_{N}\right)  ,
\label{sigmawnz}%
\end{equation}
where the reduced energy $E_{N}^{F}(z)$, $E_{N}^{\Omega}(z)$ and $\omega_{N}$
are given by%

\begin{align}
E_{N}^{F}(z)  &  =\frac{2}{1+\Lambda^{-1}}\Lambda^{(N-1)/2}E^{F}(z)\\
E_{N}^{\Omega}(z)  &  =\frac{2}{1+\Lambda^{-1}}\Lambda^{(N-1)/2}E^{\Omega
}(z)\\
\omega_{N}  &  =\frac{2}{1+\Lambda^{-1}}\Lambda^{(N-1)/2}\omega\text{.}%
\end{align}

In Eq.(\ref{iz}) an electron is destroyed in the orbital $f_{0z}$ of the left
conduction band and another electron is created in the orbital $g_{0z}$ of the
right conduction band, and vice versa, so that the application of the operator
$I(z)$ conserves the charge. In the matrices elements $\left|  \left\langle
F(z)\right|  I(z)\left|  \Omega(z)\right\rangle \right|  $ of the
Eq.(\ref{sigmawnz}) we take as initial state the fundamental state $\left|
\Omega(z)\right\rangle $, and as final states $\left|  F(z)\right\rangle $ all
the excited states whose charges are equal to the charge of the initial state.
As the energy levels of the conduction bands (left and right) are discretized,
the lines $\left|  \left\langle F(z)\right|  I(z)\left|  \Omega
(z)\right\rangle \right|  $ that appear in Eq.(\ref{sigmawnz}) are
discontinuous in $\omega_{N}$, so that rarely the final energy is equal to the
initial energy plus $\omega_{N}$ ($E_{N}^{F}(z)=E_{N}^{\Omega}(z)+\omega_{N}%
$). If the numerical approach for $\Lambda=1$ were possible, there would be
the continuous limit, where the energy levels of the conduction bands would
originate a dense spectrum for those lines. However, in this limit
($\Lambda\rightarrow1$) the computational costs would be infinite. To recover
the continuous limit the continuous parameter $z$ \cite{frota1} in the
discretization process of the conduction bands was introduced. The continuous
limit is obtained averaging in $z$ the conductivity $\sigma(\omega_{N},z)$
given by Eq.(\ref{sigmawnz}). The function $\delta\left(  \Phi(z)\right)
=\delta\left(  E_{N}^{F}(z)-E_{N}^{\Omega}(z)-\omega_{N}\right)  $ can be
written in terms of the roots $z_{i}$ of $\Phi(z)$\cite{gasiorowicz}:\bigskip%
\begin{equation}
\delta\left[  \Phi(z)\right]  =%
{\displaystyle\sum\limits_{i}}
\frac{\delta(z-z_{i})}{\left|  \dfrac{d\Phi}{dz}\right|  _{z=z_{i}}}.
\end{equation}
The average $\bar{\sigma}(\omega_{N})$ in $z$ of the function $\sigma
(\omega_{N},z)$, given by $\bar{\sigma}(\omega_{N})=\int_{z_{a}}^{z_{b}}%
\sigma(\omega_{N},z)dz/\Delta z$, with $\Delta z=z_{a}-z_{b}$, represents the
continuous spectrum of the electric conductivity, that is written as%

\begin{equation}
\bar{\sigma}(\omega_{N})=\frac{\pi}{\omega_{N}}%
{\displaystyle\sum\limits_{i}}
\frac{%
{\displaystyle\sum\limits_{F}}
\left|  \left\langle F(z_{i})\right|  I(z_{i})\left|  \Omega(z_{i}%
)\right\rangle \right|  ^{2}}{\left(  \frac{1+\Lambda^{-1}}{2}\Lambda
^{-(N-1)/2}\right)  ^{2}}\frac{1}{\left|  \dfrac{d\Phi}{dz}\right|  _{z=z_{i}%
}}. \label{sigmamedio}%
\end{equation}

\section{Results and discussion}

In this section the results of the charge transference $\tau$ as a function of
$U$, the charge average $\left\langle Q_{N}\right\rangle $ as a function of
$V_{ext}$, the quadratic charge average $\left\langle Q_{N}^{2}\right\rangle $
and the specific heat $C_{V}$ as a function of $\ $the temperature $T$, and
the electrical current $\sigma(\omega)$ are shown.

To verify the precision of the calculation, we present in Fig.(1) the results
of the charge transference $\left\langle \tau\right\rangle =2\left\langle
f_{0z}^{\dagger}g_{0z}\right\rangle $ as a function of $V_{T} $ \ (the
tunnelling matrices element), in the fundamental state, calculated by the NRG
(black circles) and the exact result obtained from the Green's function
equation of motion (full line). In the limit of $V_{T}\rightarrow0$, $\tau$ is
null, since in this limit the two conduction bands are decoupled. For large
$V_{T}$, the orbitals $f_{0z}$ and $g_{0z}$ are strongly coupled, disconnected
from the remain of the conduction bands. In this case the Hamiltonian $H_{N}$
is reduced to two decoupled conduction bands with two energy levels, one below
the base and the other above the top of the conduction bands, forming ligating
and antiligating states, $\left(  f_{oz}^{\dagger}-g_{oz}^{\dagger}\right)
\left|  0\right\rangle /\sqrt{2}$ and $\left(  f_{oz}^{\dagger}+g_{oz}%
^{\dagger}\right)  \left|  0\right\rangle /\sqrt{2}$, respectively, with
$2\left\langle f_{0z}^{\dagger}g_{0z}\right\rangle \rightarrow1$. From Fig.(1)
it can be noticed that the analytical and NRG results are in very good
agreement, with an error less than 2\%.

The results of $\left\langle \tau\right\rangle $ as a function of $U/D$
calculated by NRG (black circles) and by second order perturbation theory in
$V_{T}$ (full line), for three cases: $V_{T}=0.01D$, $V_{T}=0.1D$ and
$V_{T}=0.3D$ are shown in Fig.(2). For small $V_{T}$, the second order
perturbation theory develped in Apendices A, given by%

\begin{equation}
\left\langle \tau\right\rangle =-2V_{T}\ln\frac{U^{U}\left(  2+U\right)
^{\left(  2+U\right)  }}{\left(  1+U\right)  ^{2\left(  1+U\right)  }},
\end{equation}
is in good agreement with NRG calculation. However, for large $V_{T}$, the
perturbation theory results are precarious, mainly for small $U.$

The parameters $V_{L}$ and $V_{R}$, corresponding to the scattering potentials
of the localized charges in the mesoscopic junction, are introduced in
Fig.(3), taking $V_{T}=0.3D$ and $V_{L}=-V_{R}=0$, $0.1D$ and $0.2D$. The NRG
shows that the scattering potentials reduce the charge transference, with
more\ prominent effect for small $U$.

The NRG results for $\left\langle Q\right\rangle $ as a function of
$eV_{ext}/U$, taking $V_{L}=V_{R}=0$, $U=0.1D$ and $V_{T}=0.01D$ are shown in
Fig.(4). For zero temperature, the stairs, with steps spaced by $U/e$, is
associated with the quantization of the charge transference in the mesoscopic
junction. As $Q=e(N_{L}-N_{R})/2$, the difference between the number of
particles in the two metals assumes integer values due the effect of the
charge quantization in the tunnelling process. The thermal excitation smooths
the stairs as is shown in\ the results for k$_{B}T=0.02D$ (traced-point line)
and k$_{B}T=0.005D$ (traced line). At high temperatures \ the thermal
excitations masks the Coulomb blockade effect, reducing the spectrum of the
charge as a function of the external potential to a straight line. The
introduction of the scattering potential $V_{L}$ $=$ $-V_{R}=0.2D$ reduces the
width of the steps, due the electron-hole excitations that are created when an
electron pass from one side to the other of the mesoscopic junction. This
result suggests that the electron-hole excitations can reduces the Coulomb
blockade, by the renormalization of the parameter $U$ that, for these values
of $V_{L}$ \ and $V_{R}$, assumes the reduced value $U^{\ast}=0.881U$. For
$V_{L}$ $\neq V_{R}$, as is shown in Fig.(6), the results present an asymmetry
in the charge of the junction as a function of the \ applied external potential.

In Fig.(7) is shown the quadratic charge average $\left\langle Q^{2}%
\right\rangle $ as a function of $T$ for $V_{T}=0.1D,$ $V_{L}=V_{R}=0$ and
different values of $U$. For fixed $U$, $\left\langle Q^{2}\right\rangle $
increases with the temperature due the thermal fluctuation. Even for
temperature zero, there is $\left\langle Q^{2}\right\rangle \neq0$ originated
from the quantum fluctuations in consequence of the tunnelling from one side
to the other side of the junction. The introduction of the scattering
potentials $V_{L}$ and $V_{R}$ modify the quantum fluctuation, reflecting in
the quadratic charge average, as it is shown in Fig.(8).

In the analysis of the charge of the junction as a function of the external
potential we have observed that the Coulomb potential $U$ is reduced by the
effect of the electron-hole pairs excitation, originated from the scattering
potentials $V_{L}$ and $V_{R}$, decreasing the effect of the Coulomb blockade.
From the calculation of the specific heat it is found out that the tunnelling
matrices $V_{T}$ also renormalize the potential $U$. The specific heat is
shown in Figs.(9 and 10). As now we wish to analysis only the effect of the
tunnelling matrices $V_{T}$ on the renormalization of $U$, we will consider
the scattering potential $V_{L}=$ $V_{R}=0$. In Fig.(9) is shown the specific
heat $C_{V}/$k$_{B}$ as a function of $T$ for $V_{T}=0.01D$ and different
values of $U$, $U=0.1D$, $0.3D$ and $0.5D$. The results can be analyzed in
terms of a simple two-level system. According to the Eqs. (\ref{hq} and
\ref{q}), the electrostatic energy of the junction is $U(N_{L}-N_{R})^{2}/4$,
so that when there is an excess electron, the energy of the junction is
$\Delta=U/4$. In the limit of $V_{T}=0$, the model reduces to two decoupled
conduction bands, separated by a energy barrier $\Delta$. On the other hand,
at the other extreme, when $V_{T}$ is very large, the two conduction bands are
strongly coupled by the term $V_{T}(f_{0z}^{\dagger}g_{0z}+g_{0z}^{\dagger
}f_{0z})$, as is represented in Eq.(\ref{hn}). In this limit, the orbitals
$f_{0z}$ and $g_{0z}$ are disconnected from the two conduction bands and the
system reduces to two decoupled conduction bands, with two located states
below the base and above the top of the conduction bands, identified as
$(f_{0z}^{\dagger}\pm g_{0z}^{\dagger})/\sqrt{2}$, where the signals $-$ and
$+$ represent the ligant and anti-ligant states, respectively. Again the
system is reduced to two bands separated by an energy barrier $\Delta$. In
those two extremes, free bands and strongly coupled bands, the specific heat
is given by the specific heat of two energy levels separated by $\Delta$. In
Fig.(9) the full line represent $C_{V}/k_{B}$ of a two level system separated
by an energy $U/4$, which is in very good agreement with the NRG results for
small $V_{T}=0.01D$.

In Fig.(10) $C_{V}/k_{B}$ as a function of $T$ for a fixed $U=0.2D$ and
different values of $V_{T}=0.2D$, $0.3D$, $0.4D$, $0.5D$ and $0.6D$ are shown.
The results obtained from the NRG calculation are perfectly fitted by the
specific heat of a two level system separated by an effective Coulomb
potential $U^{\ast}/4$. The renormalized $U^{\ast}$ as a function of $V_{T} $
is presented in Fig.(11), which shows that $U^{\ast}\rightarrow U$ in the
limit of small and high $V_{T}$, corresponding to the regimes of decoupled and
strongly coupled conduction bands, respectively.

The eletrical conductivity is determined by the Kubo formula and is shown in
Fig.(12) as a function of the frequency $\omega$, for the Coulomb energy
$U=0.01D$, the scattering potential $V_{L}=V_{R}=0$, and three values of the
tunnelling matrices element $V_{T}=0.05D$, $0.10D$ and $0.15D$, with the
energy scaled by $U$ and the conductivity scaled by the conductivity $G_{0}$
for the limit of $\omega\rightarrow\infty$ \cite{frota2}, given by%

\begin{equation}
G_{0}=\frac{e^{2}}{h}\frac{\pi^{2}\left(  V_{T}/D\right)  ^{2}}{\left(
1+\left(  \pi V_{T}/2D\right)  ^{2}\right)  ^{2}}.
\end{equation}
The rate $G/G_{0}$ tends to the unit limit as $\omega/U\gg1$ and drops to zero
when $\omega/U\rightarrow0$, as is pointed out in the inset of Fig.(12). In
contrast with the classical behavior, where there is conductivity only for
$\omega/U>1$, here there is conductivity through the junction also for energy
lower then the Coulomb gap ($\omega/U<1$), due to the quantum fluctuations
that introduces uncertainty in the charge of the junction, as is shown in the
nset of Fig.(12). We also verify that the conductivity is an universal
function of $\omega/U$ (not shown in the figure), depending only on $V_{T}$.
The effect of the scattering potential $V_{L}$ and $V_{R}$ is to reduce the
conductivity, as it is shown in Fig.(13) for $V_{T}=0.1D$, $U=0.01D$ and three
different values of $V_{L}$ and $V_{R}$.

\section{Conclusion}

In this work we have studied the static and dynamic properties of a mesoscopic
junction, by using NRG to calculate the charge transference in the fundamental
state, the charge average and the quadratic charge average, the specific heat
and the electric conductivity of the junction.

For $V_{T}/D\gg1$ the orbitals $f_{0z}$ and $g_{0z}$ are decoupled from their
conduction bands, forming the states $\left(  f_{oz}^{\dagger}-g_{oz}%
^{\dagger}\right)  \left|  0\right\rangle /\sqrt{2}$ and $\left(
f_{oz}^{\dagger}+g_{oz}^{\dagger}\right)  \left|  0\right\rangle /\sqrt{2}$,
and the charge transference $2\left\langle f_{0z}^{\dagger}g_{0z}\right\rangle
\rightarrow1$. For any value of $V_{T}/D$, the charge transference is reduced
by the scattering potential $V_{L}$ and $V_{R}$. The charge average
$\left\langle Q\right\rangle $ as a function of the external potential
$V_{ext}$ presents a stairs spectrum, suggesting the charge quantization. The
width of the steps is equal to $U$ for $V_{L}$ $=$ $V_{R}=0 $, and is reduced
by the effect of the electron-hole pairs excitations for $V_{L}$ $=$
$V_{R}\neq0$. For different values of $U$, the results have shown that, even
for zero temperature, quadratic charge average $\left\langle Q^{2}%
\right\rangle \neq0$, as a consequence of the quantum fluctuations in the
junction.The calculation of the specific heat discloses that the junction can
be represented by a system of two energy levels separated $U^{\ast}$, that
depends on the tunnelling matrices element $T$. In the limit of $V_{T}=0$ or
$V_{T}\rightarrow\infty$, $U^{\ast}=U,$ since in these limits the two
conduction bands are decoupled. For intermediate values of $V_{T}$, $U^{\ast}$
varies between zero and $U$. The electrical conductivity as a function of the
frequency was obtained from the Kubo formula. The quantum fluctuations
originate electrical conductivity different of zero, even for energies lesser
then $U$, in contrast with the classical behavior, where the electrical
conductivity is zero for energies lesser then $U$. The scattering potentials
$V_{L}$ and $V_{R}$ introduce electron-hole pairs excitations near the Fermi
level, reducing the conductivity of the junction for all frequency scales. The
formalism developed in the present work can be extended for Hamiltonians that
represent other models of mesoscopic junction, as for instance metallic
islands or quantum dot, with several tunnelling canals.%

\newpage

\appendix

\section{Second order perturbation theory in $V_{T}$}

In this Appendices we have used the second order perturbation theory in
$V_{T}$ to calculate analytically the quadratic charge average and the charge
transference in the fundamental state for the Hamiltonian $H_{N}$ given by
Eq.(\ref{hn}), considering $V_{L}=V_{R}=0$. In this case the Hamiltonian
$H_{N}$ is written as%

\begin{equation}
H_{N}=H_{N}^{0}+H_{T}^{N},
\end{equation}
where%

\begin{align}
H_{N}^{0}  &  =\Lambda^{(N-1)/2}\left\{
{\displaystyle\sum_{n=0}^{N-1}}
\varepsilon_{nz}\left(  f_{nz}^{\dagger}f_{(n+1)z}+f_{(n+1)z}^{\dagger}%
f_{nz}\right)  \right. \nonumber\\
&  \left.  +%
{\displaystyle\sum_{n=0}^{N-1}}
\varepsilon_{nz}\left(  g_{nz}^{\dagger}g_{(n+1)z}+g_{(n+1)z}^{\dagger}%
g_{nz}\right)  +\overset{\thicksim}{U}Q_{N}^{2}\right\} \label{hnpert}\\
H_{T}^{N}  &  =\Lambda^{(N-1)/2}\overset{\thicksim}{V}_{T}\left(
f_{0z}^{\dagger}g_{0z}+g_{0z}^{\dagger}f_{0z}\right)  \text{,} \label{htpert}%
\end{align}
where $H_{T}^{N}$ is treated as a perturbation. The charge transference
$\left\langle \tau\right\rangle $ and the quadratic charge average
$\left\langle Q^{2}\right\rangle $ in the fundamental state are given by%

\begin{align}
\left\langle Q^{2}\right\rangle  &  =\frac{\partial}{\partial U}\left\langle
\Omega\right|  H\left|  \Omega\right\rangle \label{q2u}\\
\left\langle \tau\right\rangle  &  =\frac{1}{2}\frac{\partial}{\partial V_{T}%
}\left\langle \Omega\right|  H\left|  \Omega\right\rangle \text{,} \label{tt}%
\end{align}
where $H$ is given by Eq.(\ref{hcontinuous}).

The reduced energy of the fundamental state is
\begin{equation}
\left\langle \Omega\right|  H_{N}\left|  \Omega\right\rangle =E_{\Omega,N}%
^{0}+%
{\displaystyle\sum\limits_{F}}
\frac{\left|  \left\langle F\right|  H_{T}^{N}\left|  \Omega\right\rangle
\right|  ^{2}}{E_{F,N}^{0}-E_{\Omega,N}^{0}}\text{,} \label{hlinhan}%
\end{equation}
where $\left|  \Omega\right\rangle $ ($\left|  F\right\rangle $) is the many
particles fundamental (excited) satate of $H_{N}^{0}$, before (after) the
tunnelling event, with reduced energy$E_{\Omega,N}^{0}$ ($E_{F,N}^{0}$).
Following Wilson \cite{wilson1}, the many particles states are constructed
from many one body states, whose energy levels are given by%

\begin{align}
\eta_{\ell}  &  =\pm\Lambda^{\ell-1}\text{ \ ~\hspace*{0in}\hspace
*{0in}\hspace*{1in}for }N\text{ odd}\label{etalimpar}\\
\widehat{\eta}_{\ell}  &  =\pm\Lambda^{\ell-1/2}\text{ \hspace*{1in}for
}N\text{ even\hspace{1in} } \label{etalpar}%
\end{align}

In the fundamental sate $\left|  \Omega\right\rangle $ all the energy levels
of the two conduct bands below (above) the Fermi level are occupied\ (empty).
In the final state $\left|  F\right\rangle $, after the tunnelling, an
electron is transfered from the level $m$ of the left conduction band, below
de Fermi level, to level $\ell$ of the right conduction band, above the Fermi
level, with energy given by%

\begin{equation}
E_{F,N}^{0}=E_{\Omega,N}^{0}+\eta_{\ell}+\left|  \eta_{m}\right|
+\Lambda^{(N-1)/2}\overset{\thicksim}{U}\text{.} \label{efinal}%
\end{equation}
The operators $f_{0z}$ and $g_{0z}$ are written in terms of the operators
$a_{k}$ and $b_{k}$, which diagonalize the left and right conduction bands,
respectively, as it follows%

\begin{align}
f_{0z}  &  =\Lambda^{(N-1)/4}%
{\displaystyle\sum\limits_{k=-J}^{k=J}}
\alpha_{0k}a_{k}\label{f0zak}\\
g_{0z}  &  =\Lambda^{(N-1)/4}%
{\displaystyle\sum\limits_{k=-J}^{k=J}}
\alpha_{0k}b_{k} \label{g0zbk}%
\end{align}
where
\begin{align}
\alpha_{0k}  &  =\alpha_{0}\Lambda^{(k-1)/2}\label{alfa0k}\\
\alpha_{0}  &  =\left(  \frac{1-\Lambda^{-1}}{2}\right)  ^{1/2} \label{alfa0}%
\end{align}
and $\bigskip J=(N+1)/2$. Substituting the Eqs.(\ref{f0zak} e \ref{g0zbk} )
into the Eq.(\ref{htpert}), and the result inserted into the Eq.(\ref{hlinhan}%
), with $E_{F,N}^{0}$ given by Eq.(\ref{efinal}), we obtain%

\begin{equation}
\left\langle \Omega\right|  H_{N}\left|  \Omega\right\rangle =E_{\Omega,N}%
^{0}-2\left[  \Lambda^{(N-1)/2}\overset{\thicksim}{V}_{T}\right]  ^{2}%
{\displaystyle\sum\limits_{m=1}^{J}}
{\displaystyle\sum\limits_{\ell=1}^{J}}
\frac{\left[  \Lambda^{-(N-1)/2}\alpha_{0m}\alpha_{0\ell}\right]  ^{2}}%
{\eta_{\ell}+\eta_{m}+\Lambda^{(N-1)/2}\overset{\thicksim}{U}},
\label{hlinhan1}%
\end{equation}
To obtain $\left\langle \Omega\right|  H\left|  \Omega\right\rangle $ we
multiply both the members of the Eq.(\ref{hlinhan1}) by the factor
$\frac{1+\Lambda^{-1}}{2}\Lambda^{-(N-1)/2}$, with $\eta_{\ell}$ and
$\alpha_{0m}$ given by the Eqs.(\ref{etalimpar} and \ref{alfa0k}),
respectively. In the limit of $\Lambda\rightarrow1$, $\left\langle
\Omega\right|  H\left|  \Omega\right\rangle $ is written as%

\begin{equation}
\left\langle \Omega\right|  H\left|  \Omega\right\rangle =E_{\Omega}%
^{0}-2V_{L}^{2}\left[  \ln\frac{U^{U}\left(  2+U\right)  ^{\left(  2+U\right)
}}{\left(  1+U\right)  ^{2\left(  1+U\right)  }}\right]  \text{.}
\label{homega}%
\end{equation}

From the Eqs.(\ref{q2u}, \ref{tt} and \ref{homega}) we finally have%

\begin{align}
\left\langle Q^{2}\right\rangle  &  =-2V_{T}^{2}\ln\left[  \frac{U\left(
2+U\right)  }{\left(  1+U\right)  ^{2}}\right] \label{q2analitico}\\
\left\langle \tau\right\rangle  &  =-2V_{T}\ln\frac{U^{U}\left(  2+U\right)
^{\left(  2+U\right)  }}{\left(  1+U\right)  ^{2\left(  1+U\right)  }}\text{.}
\label{tanalitico}%
\end{align}

\strut%

\newpage

\begin{center}
FIGURE CAPTIONS
\end{center}

Figure 1. Charge transference as a function of $V_{T}/D$, for $U=V_{L}%
=V_{R}=0$. The full line represents the exact solution using the Green's
function equation of motion; the black circles are the results of NRG
calculation, considering the discretization parameter $\Lambda=3$.

\medskip Figure 2. Charge transference as a function of $U/D$, taking
$V_{L}=V_{R}=0$. The full lines represent the results of the second order
perturbation theory in $V_{T}$ and the black circles are the results obtained
from the NRG for $\Lambda=3.$

\medskip Figure 3. Charge transference as a function of $U/D$, for
$V_{T}=0.3D$, $V_{L}=-V_{R}=0$, $0.1D$ and $0.2D$, considering the
discretization parameter $\Lambda=3$.

\medskip Figure 4. The stairs bahavior of the charge of the junction as a
function of $V_{ext}$,\ for $V_{T}=0.01D$, $U=0.1D$, $V_{L}=V_{R}=0$, and
three different temperatures, given in half width of the conduction banda,
k$_{B}T/D$. The quantization of the charge and the width of the steps of the
stairs equal to $U$, which is smoothed by the temperature, are shown.

\medskip Figure 5. Charge of the junction as a function of $V_{ext}$,\ for a
symetric junction with $V_{T}=0.01D$, $U=0.1D$, the scaterring potentials
$V_{L}=-V_{R}=$ $0.2D$, and three different temperatures. The width of the
steps is $U^{\ast}=0.881U$, which is reduced by the effect of the
electron-hole pairs excitation due to the scaterring potential.

\medskip Figure 6. Charge transference as a function of $V_{ext}$ fo an
assymetric junction with $V_{L}=0.2D$, $V_{R}=-0.1D$, $V_{T}=0.01D$, $U=0.1D$
and three values of the temperature.

\medskip Figure 7. The square charge average $\left\langle Q^{2}\right\rangle
$ as a function of the temperature, for $V_{T}=0.1D$, $V_{L}=V_{R}=0$, and
several values of $U$. \ The values $\left\langle Q^{2}\right\rangle $ tend
for a finite values in the limit of $T=0$, due to the quantum fluctuations in
the fundamental state.

\medskip Figure 8. The square charge average $\left\langle Q^{2}\right\rangle
$ as a function of the temperature $T$, for $V_{T}=0.1D$, $U=0.2D$ and several
values of $V_{L}$ and $V_{R}$.

\medskip Figure 9. The specific heat as a function of $T$ without Coulomb
scattering, for small $V_{T}$ and several values of $U$. The full line
represents the specific heat for a two level system separated by $U/4.$

\medskip Figure 10. The specific heat as a function of $T$ for $U=0.2D$ and
several values of $V_{T}$. The full line represents the specific heat for a
two level system, separated by $U/4$ renormalized by the $V_{T}$ tunnelling matrice.

\medskip Figure 11. From the fitting of the NRG results for specific heat by
the results of a two-level system we have verified that the capacitor energy
$U$ is renormalized by the tunnelling matrice $V_{T}$ as is shwon in this
figure. $U^{\ast}\rightarrow U$ in the limits of $V_{T}=0$ and $V_{T}%
\rightarrow\infty$, since the two bands are decoupled in these two limits.

\medskip Figure 12. Eletrical conductivity $G$ scaled by $G_{0}$ (conductivity
for large frequence) as a function of the frequence $\omega$ scaled by $U$,
without Coulomb scattering, with $U=0.01$ and several values of $V_{T}$.

\medskip Figure 13. Eletrical conductivity $G$ scaled by $G_{0}$ (conductivity
for large frequence) as a function of the frequence $\omega$ scaled by $U$,
with Coulomb scattering, with $V_{T}=0.1D$, $U=0.01$ and several values of
$V_{L}$ and $V_{R}$.

\bigskip

\bigskip

\bigskip

\end{document}